\documentclass[11pt]{article}
\usepackage{amsfonts}
\usepackage{amssymb}
\usepackage{amsmath}
\usepackage{graphicx}

\def \Real {\mbox{$\mathbb{R}$}}

\newcommand{\ket}[1]{| #1\rangle}

\newcommand{\opbracket}[3]{\langle #1\,|\,#2\,|\,#3\rangle}

\def \qed {\hfill \rule{0.2cm}{0.2cm}\vspace{3mm}}
\newenvironment{mylist}[1]
	{\begin{list}{}{\setlength{\leftmargin}{#1}
	\setlength{\rightmargin}{0.0cm}\setlength{\labelsep}{1.3mm}
	\setlength{\labelwidth}{0.8cm}\setlength{\itemsep}{0.0cm}}}
	{\end{list}}
\newcommand{\horizontal}{
	\begin{flushleft}
	\rule{6.69in}{0.2mm}\vspace{-2mm}
	\end{flushleft}}
\newtheorem{theorem}{Theorem}
\newtheorem{lemma}{Lemma}

\newtheorem{definition}{Definition}

\begin{document}

\title{\Large\bf PSPACE has 2-round quantum interactive proof systems}

\author{John Watrous\\
	D\'epartement d'informatique et de recherche op\'erationnelle\\
	Universit\'e de Montr\'eal\\
	Montr\'eal (Qu\'ebec), Canada\\
	watrous@iro.umontreal.ca}

\maketitle

\thispagestyle{empty}

\begin{abstract}
In this paper we consider quantum interactive proof systems, i.e., interactive
proof systems in which the prover and verifier may perform quantum
computations and exchange quantum messages.
It is proved that every language in PSPACE has a quantum interactive proof
system that requires only two rounds of communication between the prover and
verifier, while having exponentially small (one-sided) probability of error.
It follows that quantum interactive proof systems are strictly more powerful
than classical interactive proof systems in the constant-round case unless the
polynomial time hierarchy collapses to the second level.
\end{abstract}


\section{Introduction}
\label{sec:introduction}

A number of recent papers have provided compelling evidence (and proof, in
some cases) that certain computational, cryptographic, and
information-theoretic tasks can be performed more efficiently by models based
on quantum physics than those based on classical physics.
For example, Shor \cite{Shor97} has shown that integers can be factored in
expected polynomial time by quantum computers, a quantum key distribution
protocol of Bennett and Brassard \cite{BennettB84} that does not rely on
intractability assumptions has been proven secure under a wide variety of
attacks \cite{BihamB+98,Mayers96,Mayers98}, and Buhrman, Cleve, and Wigderson
\cite{BuhrmanC+98} have shown various separation results between quantum and
classical two-party communication complexity models.
In this paper we introduce the quantum analogue of another
concept---interactive proof systems---and provide strong evidence that
additional power is gained by interactive proof systems in the quantum setting.

Interactive proof systems were introduced by Goldwasser, Micali, and Rackoff
\cite{GoldwasserM+89} and Babai \cite{Babai85}.
Informally, in an interactive proof system a computationally unbounded prover
interacts with a polynomial-time probabilistic verifier and attempts to
convince the verifier to accept a given input string.
A language $L$ is said to have an interactive proof system if there exists a
verifier $V$ such that (i)~there exists a prover $P$ (called an honest prover)
that can always convince $V$ to accept when the given input is in $L$, and
(ii)~no prover $P^{\prime}$ can convince $V$ to accept with nonnegligable
probability when the input is not in $L$.
The class of languages having interactive proof systems is denoted $IP$.

Based on the work of Lund, Fortnow, Karloff, and Nisan \cite{LundF+92}, Shamir
\cite{Shamir92} proved that every language in PSPACE has an interactive proof
system.
Since any language having an interactive proof system is in PSPACE
\cite{Feldman86}, this implies $\mathrm{IP} = \mathrm{PSPACE}$.
All known protocols for PSPACE require a nonconstant number of rounds of
communication between the prover and verifier, and cannot be parallelized to
require only a constant number of rounds under the assumption that the
polynomial time hierarchy is proper.
This is because the class of languages having constant-round interactive proof
systems is equivalent to the class AM \cite{Babai85,GoldwasserS89}, and hence
is contained in $\Pi_2^p$.

The main result we prove in this paper is as follows.
\begin{theorem}
\label{thm:main}
Every language in PSPACE has a 2-round quantum interactive proof system with
exponentially small probability of error.
\end{theorem}
This result contrasts with the facts mentioned above regarding classical
interactive proof systems, as it shows there are languages having 2-round
quantum interactive proof systems that do not have constant-round classical
interactive proof systems unless $\mathrm{AM} = \mathrm{PSPACE}$.

We now summarize informally our technique for proving Theorem~\ref{thm:main}.
Consider the following (unsuccessful) method for trying to reduce the number
of rounds required by a nonconstant-round protocol for PSPACE to a constant:
define the verifier so that it chooses all of its random numbers initially,
sends them all to the prover in one round (or in a constant number of rounds),
receives all the responses from the prover, and checks the validity of the
responses.
This will not work, since the prover may cheat by ``looking ahead'' and basing
its responses on random numbers that would have been sent in later rounds in
the nonconstant-round case.
However, using interactive proofs based on quantum physics, this technique
can be made to work, as the aforementioned behavior on the part of the prover
can be detected by a quantum verifier.
We now sketch the method for doing this---a formal description of the protocol
appears in Section~\ref{sec:QIPS_for_QBF}.

The prover first sends a superposition of sequences of random numbers and
corresponding responses to the verifier, and the verifier checks that the
responses are valid according to a classical protocol for PSPACE.
(It will be shown that the prover cannot cheat by giving the verifier a
superposition that is biased towards certain random sequences---the verifier
will be able to later check that the superposition is close to uniform.)
The verifier then chooses randomly one of the positions in the list of
random numbers and responses, sends back to the prover its responses starting
at this position in the list and challenges the prover to invert the
computation it performed to obtain these responses.
Let us say that the random numbers and responses up to the chosen position in
the list have {\em low-index}, and the remaining random numbers and responses
have {\em high-index}.
The low-index responses, which were not sent back to the prover in the second
round, should now depend only on the low-index random numbers (for otherwise
the prover has cheated).
The verifier may now check that the superposition of high-index random numbers
is uniform by performing an appropriately defined measurement.
However, if the prover has cheated by basing its low-index responses on
high-index random numbers, the low-index responses and high-index random
numbers will be entangled in a manner detectable by the verifier; with high
probability, the high-index random numbers will fail the uniformity test.
By performing this process itself in parallel a polynomial number of times,
the probability a cheating prover escapes detection is made exponentially
small, while the protocol still requires only two rounds of communication.

The remainder of the paper has the following organization.
In Section~\ref{sec:QIP_definition} we formally define quantum interactive
proof systems.
In Section~\ref{sec:QIPS_for_QBF} we prove Theorem~\ref{thm:main} by
presenting a 2-round quantum interactive proof system for the quantified
Boolean formula problem and proving its correctness.
We conclude with Section~\ref{sec:conclusion}, which mentions a number of
open problems regarding quantum interactive proofs.


\section{Definition of quantum interactive proof systems}
\label{sec:QIP_definition}

We now give a formal definition of quantum interactive proof systems.
We restrict our attention to constant round quantum interactive proof systems,
although the definition is easily extended to a nonconstant number of rounds.
The model for quantum computation that provides a basis for our definition
of quantum interactive proof systems is the quantum circuit model.
We will not define quantum circuits or discuss them in detail, as this has
been done elsewhere (see Yao \cite{Yao93} and Berthiaume \cite{Berthiaume97},
for example).

A $k$-round verifier $V$ is a polynomial-time computable mapping
$V:\Sigma^{\ast}\times\{0,\ldots,k\}\rightarrow\Sigma^{\ast}$, where each
$V(x,j)$ is an encoding of a quantum circuit composed of quantum gates from
some appropriately chosen universal set of gates.
Universal sets of gates/transformations have been investigated in a number of
papers \cite{AdlemanD+97,Barenco95,BarencoB+95,Deutsch89,DiVincenzo95}; for
the purposes of this paper, we will assume only that this set includes the
Walsh-Hadamard gate and any universal gate for reversible computation such
as the Fredkin gate or Toffoli gate.
Each encoding $V(x,j)$ is identified with the quantum circuit it encodes.
Since the mapping $V$ is computable in polynomial time, each circuit $V(x,j)$
must be polynomial in size.
The qubits upon which each $V(x,j)$ acts are assumed to be divided into two
groups: message qubits and ancilla qubits.
The message qubits represent the communication channel between the prover and
verifier, while the ancilla qubits represent qubits that are private to the
verifier.
One of the verifier's ancilla qubits is specified as the output qubit.

A $k$-round prover $P$ is a mapping from $\Sigma^{\ast}\times\{1,\ldots,k\}$
to the set of all quantum circuits.
No restrictions are placed on the size of each $P(x,j)$ or on the gates from
which these circuits are composed.
Similar to the case of the verifier, the qubits of the prover are divided into
message qubits and ancilla qubits.
Note that although the prover is all-powerful in a computational sense (there
is no bound on the complexity of the mapping $P$ or on the size of each
$P(x,j)$), we of course require that the prover obey the laws of physics!
This is enforced by requiring that the prover's actions correspond to quantum
circuits.

\begin{figure}[!ht]
\center{
\begin{picture}(0,0)%
\includegraphics{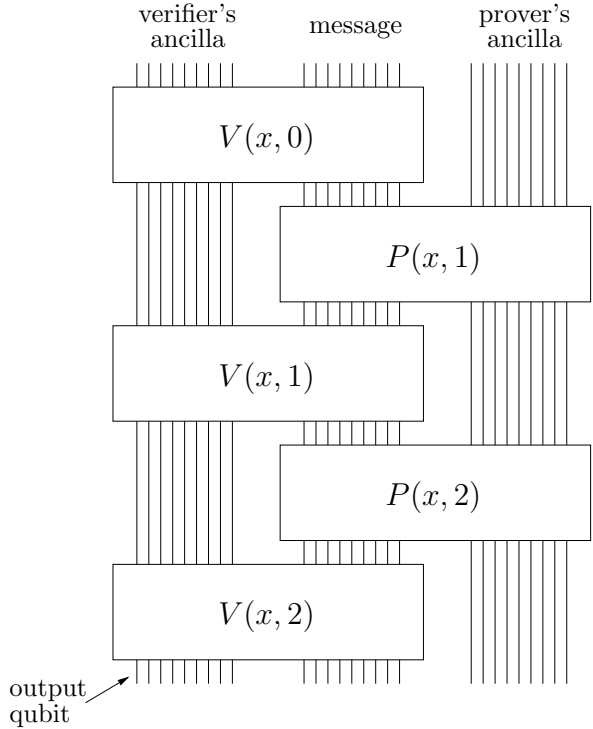}%
\end{picture}%
\setlength{\unitlength}{1973sp}%
\begingroup\makeatletter\ifx\SetFigFont\undefined
\def\x#1#2#3#4#5#6#7\relax{\def\x{#1#2#3#4#5#6}}%
\expandafter\x\fmtname xxxxxx\relax \def\y{splain}%
\ifx\x\y   
\gdef\SetFigFont#1#2#3{%
  \ifnum #1<17\tiny\else \ifnum #1<20\small\else
  \ifnum #1<24\normalsize\else \ifnum #1<29\large\else
  \ifnum #1<34\Large\else \ifnum #1<41\LARGE\else
     \huge\fi\fi\fi\fi\fi\fi
  \csname #3\endcsname}%
\else
\gdef\SetFigFont#1#2#3{\begingroup
  \count@#1\relax \ifnum 25<\count@\count@25\fi
  \def\x{\endgroup\@setsize\SetFigFont{#2pt}}%
  \expandafter\x
    \csname \romannumeral\the\count@ pt\expandafter\endcsname
    \csname @\romannumeral\the\count@ pt\endcsname
  \csname #3\endcsname}%
\fi
\fi\endgroup
\begin{picture}(7287,9102)(526,-9622)
\put(526,-9211){\makebox(0,0)[lb]{\smash{\SetFigFont{10}{12.0}{rm}output}}}
\put(526,-9541){\makebox(0,0)[lb]{\smash{\SetFigFont{10}{12.0}{rm}qubit}}}
\put(2296,-1051){\makebox(0,0)[lb]{\smash{\SetFigFont{10}{12.0}{rm}ancilla}}}
\put(2146,-736){\makebox(0,0)[lb]{\smash{\SetFigFont{10}{12.0}{rm}verifier's}}}
\put(6526,-1051){\makebox(0,0)[lb]{\smash{\SetFigFont{10}{12.0}{rm}ancilla}}}
\put(6421,-736){\makebox(0,0)[lb]{\smash{\SetFigFont{10}{12.0}{rm}prover's}}}
\put(4291,-871){\makebox(0,0)[lb]{\smash{\SetFigFont{10}{12.0}{rm}message}}}
\put(3151,-2311){\makebox(0,0)[lb]{\smash{\SetFigFont{12}{14.4}{rm}$V(x,0)$}}}
\put(3151,-5311){\makebox(0,0)[lb]{\smash{\SetFigFont{12}{14.4}{rm}$V(x,1)$}}}
\put(3151,-8311){\makebox(0,0)[lb]{\smash{\SetFigFont{12}{14.4}{rm}$V(x,2)$}}}
\put(5251,-3811){\makebox(0,0)[lb]{\smash{\SetFigFont{12}{14.4}{rm}$P(x,1)$}}}
\put(5251,-6811){\makebox(0,0)[lb]{\smash{\SetFigFont{12}{14.4}{rm}$P(x,2)$}}}
\end{picture}
}
\caption{Quantum circuit for a 2-round quantum interactive proof system}
\label{fig:QIPS}
\end{figure}

Given a pair $(P,V)$, we consider a quantum circuit composed in the manner
illustrated in Figure~\ref{fig:QIPS} (the case $k=2$ is shown).
The probability that a pair $(P,V)$ accepts a given input $x$ is defined to be
the probability that an observation of the output qubit in the
$\{\ket{0},\ket{1}\}$ basis yields $\ket{1}$ when the circuits
$V(x,0), P(x,1), V(x,1),\ldots, P(x,k), V(x,k)$ are applied in sequence as
illustrated, assuming all qubits are initially in the $\ket{0}$ state.

Now, we say that a language $L$ has a $k$-round quantum interactive proof
system with error probability $\epsilon$ if there exists a $k$-round verifier
$V$ such that
\begin{mylist}{\parindent}
\item[1.] There exists a $k$-round prover $P$ such that if $x\in L$ then
$(P,V)$ accepts $x$ with probability 1.
\item[2.] For all $k$-round provers $P^{\prime}$, if $x\not\in L$ then
$(P^{\prime},V)$ accepts $x$ with probability at most~$\epsilon$.
\end{mylist}

A few notes regarding the above definition are in order.
First, we note that there are a number of other ways in which we could have
defined quantum interactive proof systems, such as a definition based on
quantum Turing machines or a definition requiring that each circuit as above
be given by $V(|x|,i)$ or $P(|x|,i)$, with $x$ supplied as input to each
circuit, for example.
We have chosen the above definition because of its simplicity.
Given the apparent robustness of the class of ``polynomial-time computable
quantum transformations,'' we suspect these definitions to be equivalent,
although we have not investigated this question in detail.
Second, we assume that each circuit corresponds to a unitary operator (e.g.,
no ``measurement gates'' are used).
The action of any general quantum gate (i.e., a gate corresponding to a
trace-preserving, completely positive linear map on mixed states of qubits)
can always be simulated by some unitary gate (possibly adding more ancilla
qubits) \cite{AharonovK+98}.
As this will not increase the size of a verifier's circuit by more than a
polynomial factor, and will not affect the complexity of the mapping $V$
significantly, our definition is equivalent to a definition allowing more
general quantum gates.


\section{2-round quantum interactive proof systems for the QBF problem}
\label{sec:QIPS_for_QBF}

We begin this section by defining the quantified Boolean formula problem,
which is complete for PSPACE.
A quantified Boolean formula is a formula of the form
$Q_1 x_1\cdots Q_n x_n B(x_1,\ldots,x_n)$, where each $Q_i$ is an existential
or universal quantifier ($\exists$ or $\forall$) and $B(x_1,\ldots,x_n)$ is a
Boolean formula (without quantifiers) in the variables $x_1,\ldots,x_n$.
The quantified Boolean formula (QBF) problem is to determine if a quantified
Boolean formula is true.

To prove Theorem~\ref{thm:main}, it is sufficient to prove that there exists
a 2-round quantum interactive proof system with exponentially small error for
the QBF problem.
This is because a verifier (and any honest prover) may first compute a
polynomial-time reduction from a given problem in PSPACE to the QBF problem,
then execute the protocol for QBF (adjusting various parameters in the
protocol to reduce error as necessary).


\subsection{A classical protocol for QBF}
\label{sec:classical}

Our 2-round quantum interactive proof system for the QBF problem is based on a
variant of the Lund--Fortnow--Karloff--Nisan protocol due to Shen
\cite{Shen92}, to which the reader is referred for a detailed description.
In this section we review some facts regarding this protocol that will later
be helpful.

Let us suppose the input formula $Q = Q_1 x_1 \cdots Q_n x_n B(x_1,\ldots,x_n)$
is fixed.
Also let $\mathbb{F}$ be a finite field, write $N = \binom{n+1}{2}+n$, and let
$d$ be the length of $Q$ (with a slight modification of the protocol, $d=3$
is sufficient).
The protocol is as follows.
For $j = 1,\ldots, N-1$, the prover sends the verifier a polynomial $f_j$ over
$\mathbb{F}$ of degree at most $d$, and the verifier chooses
$r_j\in\mathbb{F}$ and sends $r_j$ to the prover.
The prover then sends a polynomial $f_N$ to the verifier in the final round,
and the verifier chooses $r_N\in\mathbb{F}$ (there is no need for $r_N$ to be
sent to the prover).
The verifier then evaluates a particular polynomial-time predicate
$E(Q,r_1,\ldots,r_N,f_1,\ldots,f_N)$ and accepts if and only if the predicate
evaluates to true.

A formal description of $E$ may be derived from the paper of Shen.
Since the details of the predicate are not necessary for our discussion,
we will only state certain properties of $E$.
First, for any sequence of random numbers $r_1,\ldots,r_{N}\in\mathbb{F}$
there exist polynomials $c_1,\ldots,c_N$, where each polynomial $c_j$ depends
only on $r_1,\ldots,r_{j-1}$, that correspond to the answers that should
be given by an honest prover.
These polynomials, which are well-defined regardless of the Boolean value
of $Q$, satisfy the following properties:
\begin{mylist}{6mm}
\item[1.] If $Q$ evaluates to true, then for all sequences $r_1,\ldots,r_N$,
$E(Q,r_1,\ldots,r_N,c_1,\ldots,c_N) = \mathrm{true}$.
\item[2.] If $Q$ evaluates to false, then for all sequences $r_1,\ldots,r_N$,
$E(Q,r_1,\ldots,r_N,c_1,f_2,\ldots,f_N) = \mathrm{false}$ for
all polynomials $f_2,\ldots,f_N$.
\item[3.] If $Q$ evaluates to false, then for any $k\in\{1,\ldots,N-1\}$ and
$r_1,\ldots,r_{k-1}\in\mathbb{F}$, the following holds.
If $f_1,\ldots,f_k$ are such that $f_k \not= c_k$, then there are at most $d$
values of $r_k$ for which there exist $r_{k+1},\ldots,r_N$ and
$f_{k+2},\ldots,f_N$ such that
$E(Q,r_1,\ldots,r_N,f_1,\ldots,f_k,c_{k+1},f_{k+2},\ldots,f_N)=\mathrm{true}$.
\item[4.] If $Q$ evaluates to false, then for any $r_1,\ldots,r_{N-1}$ and
$f_1,\ldots,f_N$ for which $f_N\not=c_N$, there are at most $d$ values of
$r_N$ for which $E(Q,r_1,\ldots,r_N,f_1,\ldots,f_N)=\mathrm{true}$.
\end{mylist}
For given $r_1,\ldots,r_{k-1}$, we call the polynomial $c_k$ the {\em correct}
polynomial corresponding to $r_1,\ldots,r_{k-1}$.

Clearly, if $Q$ evaluates to true, an honest prover can always convince the
verifier to accept by sending the correct polynomials $c_1,\ldots,c_N$
corresponding to the verifiers random numbers $r_1,\ldots,r_{N-1}$.

Now suppose that $Q$ evaluates to false.
By item 2, a cheating prover cannot send the correct polynomial $c_1$ on the
first round, for the prover rejects with certainty in this case.
Hence the prover must send $f_1\not=c_1$ if the verifier is to accept.
Now suppose for $k\in\{1,\ldots,N-1\}$ and $r_1,\ldots,r_{k-1}$ the prover has
sent polynomials $f_1\not=c_1,\ldots,f_k\not=c_k$ during rounds $1,\ldots,k$.
Unless the verifier randomly chooses one of $d$ particular values for $r_k$,
the prover may not send $c_{k+1}$ on the next round without causing the
verifier to reject.
Hence, if the prover sends an incorrect polynomial on round $k$, then with
probability at least $1-d/|\mathbb{F}|$ it must send an incorrect polynomial
on round $k+1$.
Finally, if the prover does not send the correct polynomial $c_N$ during the
last round, the verifier accepts with probability at most $d/|\mathbb{F}|$.
Hence, the total probability that the verifier accepts may not exceed
$(d N)/|\mathbb{F}|$.

Since the error probability of the protocol depends on the size of
$\mathbb{F}$, $\mathbb{F}$ may be chosen sufficiently large at the start of
the protocol.
It will be convenient for us to take $\mathbb{F}$ to be the field with
$2^k$ elements for $k$ polynomial in $n$ (hence yielding exponentially small
probability of error).
For any chosen $k$, the verifier (and honest prover) may use a deterministic
procedure to implement arithmetic in \mbox{$\mathbb{F}$---specifically},
compute an irreducible polynomial $g$ of degree $k$ over $GF(2)$ in
deterministic polynomial time \cite{Shoup90}, identify elements of
$\mathbb{F}$ with polynomials over $GF(2)$ of degree at most $k-1$, and take
arithmetic to be the usual arithmetic on polynomials modulo $g$.
There is thus a natural correspondence between $k$ bit strings and elements
of $\mathbb{F}$.


\subsection{Quantum verifier's protocol for QBF}

We now describe the verifier's protocol for our 2-round quantum interactive
proof system for the QBF problem.

We use the following conventions when describing the quantum circuits
corresponding to the verifier's actions.
Collections of qubits upon which various transformations are performed are
referred to as registers, and are labeled by capital letters in boldface.
The registers required by the protocol are $\mathbf{R}_{i,j}$,
$\mathbf{S}_{i,j}$, and $\mathbf{F}_{i,j}$ for $1\leq i\leq m$ and
$1\leq j\leq N$, where $N$ is as in the classical protocol described in
Section~\ref{sec:classical} and $m$ is some polynomial in $n$ specified
depending on the desired error as described below.
Each register $\mathbf{R}_{i,j}$ and $\mathbf{S}_{i,j}$ consists of $k$ qubits,
where $2^k$ is to be the size of the field $\mathbb{F}$.
We view the classical states of these registers as elements in $\mathbb{F}$
in the usual way.
Each $\mathbf{F}_{i,j}$ consists of $d+1$ collections of $k$ qubits, for $d$
as in the classical protocol, and we view the classical states of these
registers as polynomials of degree at most $d$ with coefficients in
$\mathbb{F}$.
The verifier may also use any polynomial number of additional ancilla qubits
in order to perform the transformations described.
In addition, the verifier will store the vector $u$ and any auxiliary variables
needed for the protocol---as there will be no need for the verifier to perform
quantum operations on these values, we consider them as being stored
classically (although there is no difference in the behavior of the protocol
if they are thought of as being stored in quantum registers).

The error probability of the protocol will depend on $m$ and $k$ as described
below in Section~\ref{sec:proof}---we may take $m$ and $k$ to be fixed
polynomials in $n$ to obtain exponentially small error.

\begin{figure}[!ht]
\center{
\begin{picture}(0,0)%
\includegraphics{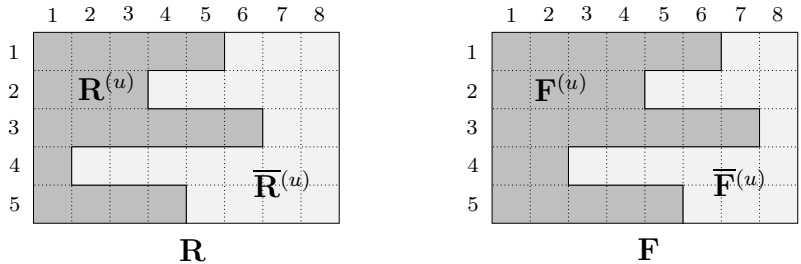}%
\end{picture}%
\setlength{\unitlength}{1579sp}%
\begingroup\makeatletter\ifx\SetFigFont\undefined
\def\x#1#2#3#4#5#6#7\relax{\def\x{#1#2#3#4#5#6}}%
\expandafter\x\fmtname xxxxxx\relax \def\y{splain}%
\ifx\x\y   
\gdef\SetFigFont#1#2#3{%
  \ifnum #1<17\tiny\else \ifnum #1<20\small\else
  \ifnum #1<24\normalsize\else \ifnum #1<29\large\else
  \ifnum #1<34\Large\else \ifnum #1<41\LARGE\else
     \huge\fi\fi\fi\fi\fi\fi
  \csname #3\endcsname}%
\else
\gdef\SetFigFont#1#2#3{\begingroup
  \count@#1\relax \ifnum 25<\count@\count@25\fi
  \def\x{\endgroup\@setsize\SetFigFont{#2pt}}%
  \expandafter\x
    \csname \romannumeral\the\count@ pt\expandafter\endcsname
    \csname @\romannumeral\the\count@ pt\endcsname
  \csname #3\endcsname}%
\fi
\fi\endgroup
\begin{picture}(12387,3990)(2026,-3946)
\put(2641,-196){\makebox(0,0)[lb]{\smash{\SetFigFont{8}{9.6}{rm}1}}}
\put(3226,-181){\makebox(0,0)[lb]{\smash{\SetFigFont{8}{9.6}{rm}2}}}
\put(3826,-166){\makebox(0,0)[lb]{\smash{\SetFigFont{8}{9.6}{rm}3}}}
\put(4411,-181){\makebox(0,0)[lb]{\smash{\SetFigFont{8}{9.6}{rm}4}}}
\put(5041,-181){\makebox(0,0)[lb]{\smash{\SetFigFont{8}{9.6}{rm}5}}}
\put(5626,-181){\makebox(0,0)[lb]{\smash{\SetFigFont{8}{9.6}{rm}6}}}
\put(6241,-181){\makebox(0,0)[lb]{\smash{\SetFigFont{8}{9.6}{rm}7}}}
\put(6826,-181){\makebox(0,0)[lb]{\smash{\SetFigFont{8}{9.6}{rm}8}}}
\put(9841,-196){\makebox(0,0)[lb]{\smash{\SetFigFont{8}{9.6}{rm}1}}}
\put(10426,-181){\makebox(0,0)[lb]{\smash{\SetFigFont{8}{9.6}{rm}2}}}
\put(11026,-166){\makebox(0,0)[lb]{\smash{\SetFigFont{8}{9.6}{rm}3}}}
\put(11611,-181){\makebox(0,0)[lb]{\smash{\SetFigFont{8}{9.6}{rm}4}}}
\put(12241,-181){\makebox(0,0)[lb]{\smash{\SetFigFont{8}{9.6}{rm}5}}}
\put(12826,-181){\makebox(0,0)[lb]{\smash{\SetFigFont{8}{9.6}{rm}6}}}
\put(13441,-181){\makebox(0,0)[lb]{\smash{\SetFigFont{8}{9.6}{rm}7}}}
\put(14026,-181){\makebox(0,0)[lb]{\smash{\SetFigFont{8}{9.6}{rm}8}}}
\put(2026,-766){\makebox(0,0)[lb]{\smash{\SetFigFont{8}{9.6}{rm}1}}}
\put(2026,-1381){\makebox(0,0)[lb]{\smash{\SetFigFont{8}{9.6}{rm}2}}}
\put(2026,-1951){\makebox(0,0)[lb]{\smash{\SetFigFont{8}{9.6}{rm}3}}}
\put(2041,-2551){\makebox(0,0)[lb]{\smash{\SetFigFont{8}{9.6}{rm}4}}}
\put(2041,-3166){\makebox(0,0)[lb]{\smash{\SetFigFont{8}{9.6}{rm}5}}}
\put(9226,-766){\makebox(0,0)[lb]{\smash{\SetFigFont{8}{9.6}{rm}1}}}
\put(9226,-1381){\makebox(0,0)[lb]{\smash{\SetFigFont{8}{9.6}{rm}2}}}
\put(9226,-1951){\makebox(0,0)[lb]{\smash{\SetFigFont{8}{9.6}{rm}3}}}
\put(9241,-2551){\makebox(0,0)[lb]{\smash{\SetFigFont{8}{9.6}{rm}4}}}
\put(9241,-3166){\makebox(0,0)[lb]{\smash{\SetFigFont{8}{9.6}{rm}5}}}
\put(4711,-3946){\makebox(0,0)[lb]{\smash{\SetFigFont{12}{14.4}{rm}$\mathbf{R}$}}}
\put(11911,-3931){\makebox(0,0)[lb]{\smash{\SetFigFont{12}{14.4}{rm}$\mathbf{F}$}}}
\put(5881,-2941){\makebox(0,0)[lb]{\smash{\SetFigFont{12}{14.4}{rm}$\overline{\mathbf R}{}^{(u)}$}}}
\put(13096,-2926){\makebox(0,0)[lb]{\smash{\SetFigFont{12}{14.4}{rm}$\overline{\mathbf F}{}^{(u)}$}}}
\put(3121,-1411){\makebox(0,0)[lb]{\smash{\SetFigFont{12}{14.4}{rm}$\mathbf{R}^{(u)}$}}}
\put(10291,-1426){\makebox(0,0)[lb]{\smash{\SetFigFont{12}{14.4}{rm}$\mathbf{F}^{(u)}$}}}
\end{picture}
}
\caption{Example division of $\mathbf{R}$ and $\mathbf{F}$ for $N = 8$,
$m = 5$, and $u = (6,4,7,2,5)$.}
\label{fig:registers}
\end{figure}

It will be convenient to refer to certain collections of the quantum registers
mentioned above; for a given vector $u\in\{1,\ldots,N\}^{m}$ we let
$\mathbf{R}^{(u)}$ be the collection of registers
$\mathbf{R}_{i,1},\ldots,\mathbf{R}_{i,u_{i}-1}$ for $i=1,\ldots,m$, and we
let $\mathbf{F}^{(u)}$ be the collection of registers
$\mathbf{F}_{i,1},\ldots,\mathbf{F}_{i,u_{i}}$ for $i=1,\ldots,m$.
See Figure~\ref{fig:registers} for an example.
We also let $\mathbf{R}_i$ and $\mathbf{F}_i$ denote the vectors
$(\mathbf{R}_{i,1},\ldots,\mathbf{R}_{i,N})$ and
$(\mathbf{F}_{i,1},\ldots,\mathbf{F}_{i,N})$, respectively.

The verifier's protocol is described in Figure~\ref{fig:QIPS_for_QBF}.
\begin{figure}[!ht]
\horizontal
\begin{mylist}{6mm}
\item[1.] 
	Receive quantum registers $\mathbf R$ and $\mathbf F$ from the prover.
	Reject if $({\mathbf R}_i,{\mathbf F}_i)$ contain an invalid proof
	that the input formula $Q$ evaluates to true for any
	$i\in\{1,\ldots,m\}$.
\item[2.]
	Choose $u\in\{1,\ldots,N\}^m$ uniformly at random and send $u$ and
	$\overline{\mathbf F}{}^{(u)}$ to the prover.
\item[3.]
	Receive $\mathbf S$ from the prover and subtract $\mathbf R_{i,j}$ from
	$\mathbf S_{i,j}$ for each $i,j$.
\item[4.]
	Apply transformation $H^{\otimes k}$ to each register of
	$\overline{\mathbf R}{}^{(u)}$.
	If $\overline{\mathbf R}{}^{(u)}$ now contains only 0 values, then
	accept, otherwise reject.\vspace{-3mm}
\end{mylist}
\horizontal
\caption{Quantum verifier's protocol for the QBF problem.}
\label{fig:QIPS_for_QBF}
\end{figure}
The check in step 1 refers to the classical protocol described in
Section~\ref{sec:classical}.
Naturally this check is performed by reversibly computing the predicate $E$
(described in Section~\ref{sec:classical}), so as not to alter superpositions
of valid pairs $(R,F)$.
The transformation $H^{\otimes k}$ in step 4 is the Walsh-Hadamard transform
applied to each qubit of the register in question, where
\[
H:\ket{0}\mapsto\frac{1}{\sqrt{2}}(\ket{0}+\ket{1})
\rule{1cm}{0cm}\mathrm{and}\rule{1cm}{0cm}
H:\ket{1}\mapsto\frac{1}{\sqrt{2}}(\ket{0}-\ket{1})
\]
as usual.
The random choice of the vector $u$ in step 2 can be simulated efficiently
with negligible error using the Walsh-Hadamard transform appropriately.
(Note that this negligible error will not change the fact that the protocol
has one-sided error.)


\subsection{Proof of correctness}
\label{sec:proof}

We now prove that the above protocol is correct.
First we show that there exists an honest prover $P$ such that $(P,V)$
accepts with certainty whenever the input formula $Q$ evaluates to true.

Given QBF $Q$ and $m\times N$ matrix $R$ of elements in $\mathbb{F}$, let
$C(R)$ denote the corresponding matrix of correct polynomials as defined in
Section~\ref{sec:classical}.
For each $i$, $C(R)_{i,1},\ldots,C(R)_{i,N}$ is thus the sequence of
polynomials the honest prover returns in the classical protocol given random
numbers $R_{i,1},\ldots,R_{i,N}$.
The honest (quantum) prover first prepares superposition
\[
2^{-kmN/2}\sum_{R}\ket{R}\ket{C(R)}
\]
in registers $\mathbf R$ and $\mathbf F$, adds the contents of each register
${\mathbf R}_{i,j}$ to ${\mathbf S}_{i,j}$, and sends $\mathbf R$ and
$\mathbf F$ to the verifier.
Under the assumption $Q$ is true, each pair $(R_i,F_i)$ the verifier receives
is a valid pair with respect to the classical protocol, so the verifier will
not reject in step 1.

The behavior of the honest prover in the second round is as follows.
For each $i,j$, let $T_{i,j}$ be a unitary transformation such that
\[
T_{i,j}:\ket{R}\ket{0}\mapsto\ket{R}\ket{C(R)_{i,j}}.
\]
Upon receiving $u$ and $\overline{\mathbf F}{}^{(u)}$ in the second round, the
prover applies transformation $T^{-1}_{i,j}$ to $\mathbf{S}$ together with
${\mathbf F}_{i,j}$ for each appropriate pair $i,j$.
This returns each register of $\overline{\mathbf F}{}^{(u)}$ to its initial
zero value.
The prover then sends $\mathbf{S}$ to the verifier.
It may be checked that after subtracting each ${\mathbf R}_{i,j}$ from
${\mathbf S}_{i,j}$, the registers $\overline{\mathbf R}{}^{(u)}$ will not be
entangled with any other registers (as each register of ${\mathbf F}{}^{(u)}$
depends only on those of ${\mathbf R}{}^{(u)}$), and are in a uniform
superposition over all possible values.
Thus, each register of $\overline{\mathbf R}{}^{(u)}$ is put into state 0
during step 4, and hence the verifier accepts with certainty.

Now we show that the verifier accepts with exponentially small probability
in case $Q$ is false, given any prover.
We begin by examining the total state of the prover and verifier as the
protocol is executed.
In step 1 the prover sends registers $\mathbf R$ and $\mathbf F$ to the
verifier.
The state of the system at this point may be expressed as
\[
\ket{\psi}\:=\:\sum_{R,F}\alpha(R,F)\ket{R}\ket{F}\ket{\xi(R,F)},
\]
where each $\alpha(R,F)$ is a complex number and $\ket{\xi(R,F)}$ is a
normalized vector representing the state of the prover's ancilla registers
(which may be entangled with $\mathbf R$ and $\mathbf F$ in any manner the
prover chooses).
Since the verifier rejects any pair $R,F$ for which each $(R_i,F_i)$ is not a
valid proof that $Q$ is true, we may assume $\ket{\psi}$ is a superposition
over such valid pairs for the purposes of bounding the probability that the
verifier accepts.

At this point, let us associate with each register $\mathbf{R}_{i,j}$ and each
register $\mathbf{F}_{i,j}$ a random variable.
The probabilities with which each random variable takes a particular value is
precisely the probability that an observation of the associated register
yields the given value, assuming that the observation takes place while the
entire system is in state $\ket{\psi}$ above.
As we have done above for registers, we may consider collections of random
variables as being single random variables, abbreviated by $\mathbf{R}^{(u)}$,
$\mathbf{F}^{(u)}$, etc.
For example,
\[
\mathrm{Pr}[\mathbf{R} = R,\,\mathbf{F}^{(u)} = F^{(u)}] \:=\:
\left\|\sum_{\overline{F}{}^{(u)}}
\alpha(R,F)\ket{\overline{F}{}^{(u)}}\ket{\xi(R,F)}\right\|^2.
\]
We also define a number of events based on these random variables.
Recall the definition of $C(R)$ from above (i.e., $C(R)$ is the $m\times N$
matrix of correct polynomials an honest prover answers for given $R$).
For $1\leq i\leq m$ and $1\leq j\leq N-1$, define $A_{i,j}$ to be the event
that $\mathbf{F}_{i,j^{\prime}}$ does not contain $C(R)_{i,j{^\prime}}$ for
$j^{\prime}\leq j$ and $\mathbf{F}_{i,j+1}$ does contain $C(R)_{i,j+1}$, for
$R$ denoting the contents of $\mathbf R$.
For $1\leq i\leq m$, define $A_{i,N}$ to be the event that
$\mathbf{F}_{i,j^{\prime}}$ does not contain $C(R)_{i,j{^\prime}}$ for every
$j^{\prime}$.
Note that we must have $\mathrm{Pr}[A_{i,1}\cup\cdots\cup A_{i,N}] = 1$ for
each $i$, as the verifier surely rejects in step 1 if $\mathbf{F}_{i,1}$
contains $C(R)_{i,1}$.
Finally, for each $v\in\{1,\ldots,N\}^m$ define events $B_v$ and $D_v$ as
$B_v = \bigcup_{i}A_{i,v_i}$ and $D_v = \bigcap_{i}A_{i,v_i}$.

In step 2 the verifier chooses $u$ randomly and sends $u$ and
$\overline{\mathbf F}{}^{(u)}$ to the prover.
The prover applies some transformation to its registers (now including
$\overline{\mathbf F}{}^{(u)}$), sends some register $\mathbf S$ to the
verifier, and the verifier subtracts the contents of $\mathbf R$ from
$\mathbf S$.
The state of the system may now be described by
\[
\sum_{R,F^{(u)}}\beta(R,u,F^{(u)})\ket{R}\ket{F^{(u)}}\ket{\eta(R,u,F^{(u)})},
\]
where each $\beta(R,u,F^{(u)})$ is a complex number and
 $\ket{\eta(R,u,F^{(u)})}$ is a normalized vector describing the state of the
prover's registers as well as register $\mathbf S$.

The verifier now executes step 4.
Assuming for now that $u$ is fixed, this results in acceptance with probability
\begin{eqnarray*}
\lefteqn{\left\|\sum_{R,F^{(u)}}\beta(R,u,F^{(u)})\ket{R^{(u)}}
\opbracket{0}{H^{\otimes k}}{R_{1,u_1}}\cdots\opbracket{0}{H^{\otimes k}}
{R_{m,u_m}}\ket{F^{(u)}}\ket{\eta(R,u,F^{(u)})}\right\|^2}\hspace{6cm}\\
&=& 2^{-lk}\sum_{R^{(u)},F^{(u)}}\left\|\sum_{\overline{R}{}^{(u)}}
\beta(R,u,F^{(u)})\ket{\eta(R,u,F^{(u)})}\right\|^2,
\end{eqnarray*}
where $l$ denotes the number of registers to which $H^{\otimes k}$ was applied,
i.e., $l = \sum_{i=1}^{m}(n - u_i + 1)$.
By the triangle inequality, this probability is at most
\begin{equation}
\label{eq:prob_bound1}
2^{-lk}\sum_{R^{(u)},F^{(u)}}\left(\sum_{\overline{R}{}^{(u)}}
\left|\beta(R,u,F^{(u)})\right|\right)^2.
\end{equation}

We now derive an upper bound on (\ref{eq:prob_bound1}) by considering the
random variables defined above.
First, we state a definition and prove a lemma regarding this definition that
will be useful for this task.

\begin{definition}
\label{def:theta}
For any nonempty, finite set $S$ and mapping $f:S\rightarrow\Real^{+}$, define
\[
\theta_{S}(f) \: = \: \frac{1}{|S|}\left(\sum_{s\in S}\sqrt{f(s)}\right)^2.
\]
\end{definition}

\begin{lemma}
\label{lemma:theta}
Let $f,g:S\rightarrow\Real^{+}$ satisfy $\sum_{s\in S}f(s)\leq 1$ and
$\sum_{s\in S}g(s)\leq 1$, let $\lambda\in[0,1]$, and let
$r = |\{s\in S|f(s) = 0\}|/|S|$.
Then
\[
\theta_{S}(\lambda f+(1-\lambda) g) \:\leq\: 1-\lambda r+2\sqrt{1-r}.
\]
\end{lemma}
{\bf Proof.}
First note that for any set $T\subseteq S$ and function
$h:T\rightarrow\Real^{+}$ with $\sum_{s\in T}h(s)\leq 1$, we have
\[
\sum_{s\in T}\sqrt{h(s)} \:\leq\: \sqrt{|T|}\,\sqrt{\sum_{s\in T}h(s)}
\]
by the Cauchy-Schwarz inequality, and hence $\theta_T(h)\leq 1$.
Now define $S^{\prime} = \{s\in S|f(s) = 0\}$.
We have
\begin{eqnarray*}
\sqrt{\theta_S(\lambda f + (1-\lambda) g)} & = &
\frac{1}{\sqrt{|S|}}\sum_{s\in S}\sqrt{\lambda f(s) + (1-\lambda)g(s)}\\
& = & \frac{\sqrt{(1-\lambda)r}}{\sqrt{|S^{\prime}|}}
\sum_{s\in S^{\prime}}\sqrt{g(s)}
+ \frac{\sqrt{1-r}}{\sqrt{|S\backslash S^{\prime}|}}
\sum_{s\in S\backslash S^{\prime}}\sqrt{\lambda f(s) + (1-\lambda)g(s)}\\
& = &  \sqrt{(1-\lambda) r}\sqrt{\theta_{S^{\prime}}(g)} +
\sqrt{1-r}\sqrt{\theta_{S\backslash S^{\prime}}(\lambda f+(1-\lambda) g)}\\
& \leq & \sqrt{(1-\lambda) r} + \sqrt{1-r}.
\end{eqnarray*}
Thus
$\theta_S(\lambda f+(1-\lambda) g)\:\leq\:1-\lambda r+2\sqrt{(1-\lambda) r (1-r)}
\:\leq\: 1-\lambda r+2\sqrt{1-r}$
as claimed.
\qed

Now, note that
\[
\left|\beta(R,u,F^{(u)})\right|^2 \:=\:
\mathrm{Pr}\left[\mathbf{R} = R,\,\mathbf{F}^{(u)} = F^{(u)}\right]
\]
for each $R$ and $F^{(u)}$; the actions of the prover and verifier are
norm-preserving, and hence will not affect the probabilities corresponding to
each $R$ and $F^{(u)}$.
Thus (\ref{eq:prob_bound1}) may be rewritten
\begin{equation}
\label{eq:prob_bound2}
2^{-lk}\sum_{R^{(u)},F^{(u)}}\left(\sum_{\overline{R}{}^{(u)}}
\sqrt{\mathrm{Pr}\left[\mathbf{R} = R,\mathbf{F}^{(u)} = F^{(u)}\right]}
\right)^2.
\end{equation}
For each pair $R^{(u)},F^{(u)}$, define a mapping
$X_{R^{(u)},F^{(u)}}:\mathbb{F}^{\,l}\rightarrow[0,1]$ as follows:
\[
X_{R^{(u)},F^{(u)}}\left(\overline{R}{}^{(u)}\right) \:=\:
\mathrm{Pr}\left[\overline{\mathbf{R}}{}^{(u)} = \overline{R}{}^{(u)}
\left|\mathbf{R}^{(u)} = R^{(u)},\mathbf{F}^{(u)} = F^{(u)}\right.\right].
\]
The probability in (\ref{eq:prob_bound2}) may be written as
\begin{equation}
\label{eq:prob_bound3}
\sum_{R^{(u)},F^{(u)}}\mathrm{Pr}\left[\mathbf{R}^{(u)} = R^{(u)},
\mathbf{F}^{(u)} = F^{(u)}\right]\theta_{\mathbb{F}^{\,l}}
(X_{R^{(u)},F^{(u)}}).
\end{equation}
Define $Y_{R^{(u)},F^{(u)}}:\mathbb{F}^{\,l}\rightarrow[0,1]$ and
$Z_{R^{(u)},F^{(u)}}:\mathbb{F}^{\,l}\rightarrow[0,1]$ as follows:
\begin{eqnarray*}
Y_{R^{(u)},F^{(u)}}\left(\overline{R}{}^{(u)}\right) & = &
\mathrm{Pr}\left[\left.\overline{\mathbf{R}}{}^{(u)}=
\overline{R}{}^{(u)}\right|\mathbf{R}^{(u)}=R^{(u)},\,\mathbf{F}^{(u)}=
F^{(u)},\,B_u\right],\\[2mm]
Z_{R^{(u)},F^{(u)}}\left(\overline{R}{}^{(u)}\right) & = &
\mathrm{Pr}\left[\left.\overline{\mathbf{R}}{}^{(u)}=
\overline{R}{}^{(u)}\right|\mathbf{R}^{(u)}=R^{(u)},\,\mathbf{F}^{(u)}=
F^{(u)},\,\neg B_u\right],
\end{eqnarray*}
for events $B_u$ and $\neg B_u$ defined previously.
We have
\[
\theta_{\mathbb{F}^{\,l}}(X_{R^{(u)},F^{(u)}})
\:=\:\theta_{\mathbb{F}^{\,l}}\left(\lambda_{u}\,Y_{R^{(u)},F^{(u)}}
+(1-\lambda_{u})\,Z_{R^{(u)},F^{(u)}}\right).
\]
for
$\lambda_u = \mathrm{Pr}\left[B_u\left|\mathbf{R}^{(u)}=R^{(u)},
\mathbf{F}^{(u)}=F^{(u)}\right.\right]$.

Now consider the values of $\overline{R}{}^{(u)}$ for which
$Y_{R^{(u)},F^{(u)}}\left(\overline{R}{}^{(u)}\right)=0$; we claim the number
of such values is at least $\left(1-dm2^{-k}\right)\,2^{kl}$ for every
$R^{(u)},F^{(u)}$.
This may be argued as follows.
First, fix values for $R^{(u)}$, $F^{(u)}$, and $i$, and assume event
$A_{i,u_i}$ takes place.
By the properties of the classical protocol discussed in
Section~\ref{sec:classical}, there are at most $d$ values of $R_{i,u_i}$
that do not cause the classical protocol to reject in this case.
Thus, the number of values of $\overline{R}{}^{(u)}$ for which
\[
\mathrm{Pr}\left[\left.\overline{\mathbf{R}}{}^{(u)}=
\overline{R}{}^{(u)}\right|\mathbf{R}^{(u)}=R^{(u)},\,\mathbf{F}^{(u)}=
F^{(u)},\,A_{i,u_i}\right]\not=0
\]
is at most $d\,2^{k(l-1)}$.
Since we have
\[
0 \:\leq\: Y_{R^{(u)},F^{(u)}}\left(\overline{R}{}^{(u)}\right) \:\leq\:
\sum_{i=1}^{m}\mathrm{Pr}\left[\left.\overline{\mathbf{R}}{}^{(u)}=
\overline{R}{}^{(u)}\right|\mathbf{R}^{(u)}=R^{(u)},\,\mathbf{F}^{(u)}=
F^{(u)},\,A_{i,u_i}\right],
\]
the total number of values of $\overline{R}{}^{(u)}$ for which
$Y_{R^{(u)},F^{(u)}}\left(\overline{R}{}^{(u)}\right)\not=0$ is at most
$d m 2^{k(l-1)}$.

Now we may apply Lemma~\ref{lemma:theta} to obtain
\[
\theta_{\mathbb{F}^{\,l}}\left(X_{R^{(u)},F^{(u)}}\right) \:\leq\:
1-\mathrm{Pr}\left[B_u\left|\mathbf{R}^{(u)}=R^{(u)},
\mathbf{F}^{(u)}=F^{(u)}\right.\right]\,\left(1-dm2^{-k}\right)
+ 2\sqrt{dm2^{-k}},
\]
and hence
\begin{equation}
\sum_{R^{(u)},F^{(u)}}\mathrm{Pr}\left[\mathbf{R}^{(u)} = R^{(u)},
\mathbf{F}^{(u)}=F^{(u)}\right]\theta_{\mathbb{F}^{\,l}}
\left(X_{R^{(u)},F^{(u)}}\right) \:\leq\: 
1-\mathrm{Pr}[B_u]\,\left(1-dm2^{-k}\right) + 2\sqrt{dm2^{-k}}.
\label{eq:prob_bound4}
\end{equation}

It remains to bound (\ref{eq:prob_bound4}), given that $u$ is chosen uniformly
from $\{1,\ldots,N\}^m$.
Let $U$ denote the random variable corresponding to the verifier's choice of
$u$.
We bound $\mathrm{Pr}[B_U]$ by conditioning on the events $D_v$ that describe
the exact places where the prover tries to ``sneak in'' the correct
polynomials.
Specifically, we have
\begin{multline*}
\mathrm{Pr}[B_U] \:=\: \sum_u\mathrm{Pr}[B_u]\,\mathrm{Pr}[U = u] \: = \:
N^{-m}\sum_{u,v}\mathrm{Pr}[B_u|D_v]\,\mathrm{Pr}[D_v] \\
= \: N^{-m}\sum_v\left(N^m - (N-1)^m\right)\,\mathrm{Pr}[D_v]
\: = \: 1 - \left(1 - \frac{1}{N}\right)^m \:>\: 1 - e^{-m/N}.
\end{multline*}
Thus, the overall probability that the verifier accepts is at most
\[
1-\left(1 - e^{-m/N}\right)\,\left(1-dm2^{-k}\right) + 2\sqrt{dm2^{-k}}.
\]
By initially choosing $m$ and $k$ to be sufficiently fast growing polynomials
in the input size $|x|$ (e.g., $m = (|x|+1)N$ and
$k = 2|x| + 6 + \lceil\log (d m)\rceil$), this probability
may be made smaller than $2^{-|x|}$, which completes the proof.


\section{Conclusions and Open Problems}
\label{sec:conclusion}

We have defined in this paper a natural quantum analogue of the notion of an
interactive proof system, and proved that there exist 2-round quantum
interactive proof systems with exponentially small error for any PSPACE
language.
We do not know if constant-round quantum interactive proofs characterize
PSPACE, or if there are such proof systems for (presumably) larger classes
(e.g., does NEXP have constant-round quantum interactive proofs?).
We have investigated neither the polynomial round case nor the $k$-round case
for $k>2$; what languages have such quantum proof systems?

Several variants on interactive proof systems have been studied, such as
multiprover interactive proofs \cite{BabaiF+91,Ben-OrG+88,CaiC+94,FeigeL92,
FortnowR+94}, probabilistically checkable proofs \cite{AroraS98,FortnowR+94},
and interactive proof systems having verifiers with very limited computing
power \cite{CondonL95,DworkS92}.
How do quantum analogues of these models compare with their classical
counterparts?


\bibliographystyle{plain}


\end{document}